\documentclass[useAMS,usenatbib,letterpaper]{mnras}
\usepackage{epsfig,amsmath,amssymb,natbib}
\usepackage{aas_macros}

\hypersetup{draft}
\begin{document}

\title[Var/Cov of CIC-probabilities]{\LARGE The Variance and Covariance of Counts-in-Cells Probabilities}

\author[A. Repp \& I. Szapudi]{Andrew Repp\ \& Istv\'an Szapudi\\Institute for Astronomy, University of Hawaii, 2680 Woodlawn Drive, Honolulu, HI 96822, USA}

\date{\today}

\label{firstpage}
\pagerange{\pageref{firstpage}--\pageref{lastpage}}
\maketitle

\begin{abstract}
Counts-in-cells (CIC) measurements contain a wealth of cosmological information yet are seldom used to constrain theories. Although we can predict the shape of the distribution for a given cosmology,  to fit a model to the observed CIC probabilities requires the covariance matrix -- both the variance of counts in one probability bin and the covariance between counts in different bins. To date, there have been no general expressions for these variances. Here we show that correlations of particular levels, or ``slices,''  of the density field determine the variance and covariance of CIC probabilities. We derive explicit formulae that accurately predict the variance and covariance among subvolumes of a simulated galaxy catalog, opening the door to the use of CIC measurements for cosmological parameter estimation.
\end{abstract}


\section{Introduction}

One of the primary motivations for galaxy surveys is their ability to constrain cosmological models, a consequence of the statistical imprint which cosmology leaves upon the galaxy distribution. The galaxy power spectrum is an observable which captures a large amount of the information inherent in these surveys (e.g., \citealp{Peebles1980, BaumgartFry1991}); indeed, for a Gaussian field the power spectrum encodes \emph{all} of the field's cosmologically relevant information.

However, the power spectrum is sensitive to the second moment of the distribution only. Thus, since the matter distribution is non-Gaussian (notably so on scales of $10h^{-1}$Mpc or less), the power spectrum is blind to the cosmological information residing in the higher moments on these scales \citep{MeiksinWhite1999, RimesHamilton2005}. Even worse, because the distribution is close to lognormal, a significant amount of information escapes the entire hierarchy of $N$-point correlation functions \citep{Carron2011, CarronNeyrinck2012}. Other statistical tools are thus necessary to capture the information lost in the power spectrum.

The log transform -- again because of the approximate lognormality of the distribution -- represents one means of recapturing this information \citep{Neyrinck2006, NSS09, Repp2017}; indeed, the log power spectrum captures virtually all of it \citep{CarronSzapudi2013, CarronSzapudi2014}. Another means of recapturing at least some of this information is to consider the (one-point) counts-in-cells (CIC) probability distribution function (PDF). Although this measure ignores the information inherent in spatial correlations, its higher-order moments encode information to which the power spectrum is blind, and thus joint analysis of the PDF and power spectrum can provide significantly tighter constraints on cosmological parameters than analysis of the power spectrum alone \citep{Uhlemann2019}.

Theoretical work on cosmological applications of CIC date at least to the efforts of \citet{BalianSchaeffer1989} -- who derive the form of the matter PDF under the assumption of scale-invariant $N$-point correlation functions -- resulting in a model by \citet{BernardeauSchaeffer1991} for galaxy multiplicity functions. Analysis continued with the study by \citet{Colombi1994} of log moments (via the Edgeworth expansion); the proposal by \citet{BernardeauKofman1995} of methods for generating PDFs; and the derivation by \citet{Bernardeau1994b,Bernardeau1994a} of cumulants for the matter PDF. \citet{Colombi1995} in turn examine the errors introduced by finite-volume effects, and \citet{SzapudiColombi1996} characterize the effects of cosmic-variance error. Other existing theoretical work includes analysis of sampling effects \citep{Colombi1998} and of the distribution of probability measurements \citep{Szapudi2000}. \citet{Valageas2002} provides a non-perturbative calculation of the PDF in the quasilinear regime; likewise, \citet{Uhlemann2018..473, Uhlemann2018..477, Uhlemann2019} use large deviation statistics to predict the CIC for galaxy surveys.

CIC measurements also have a long history, including their use in simulations by \citet{Baugh1995} to determine the $N$-point correlation functions and by \citet{Colombi2000} to determine the void probability distribution. Application to survey data includes analyses by \citet{SzapudiEtal1992}, \citet{Gaztanaga1994}, and \citet{SzapudiEtal1995,SzapudiEtal1996} of early projected surveys, and by \citet{Baugh2004} and \citet{Croton2007} of the 2dFGRS data; measurement by \citet{PapaiSzapudi2010} of the PDF in the SDSS LRG sample; determination by \citet{Wolk2013} of higher-order statistics in the CFTHLS-W survey; and verification by \citet{Clerkin2017} of log-normality of the projected CIC in DES data.

Turning to actual constraints on cosmological parameters, \citet{Gruen2018}, together with \citet{Friedrich2018}, provide the first complete cosmological analysis of the galaxy density PDF (in combination with lensing), thereby deriving cosmological constraints from DES and SDSS data. In addition, \citet{Salvador2019} use CIC statistics to analyze nonlinear galaxy bias in DES data, and \citet{ReppSDSSLin} derive joint constraints on $\sigma_8$ and linear galaxy bias from CIC in SDSS data.
 
However, any use of CIC measurements to constrain cosmology requires an accounting for both the variance in the probability measurements and the covariance between measurements in different probability bins. It is likely that this fact plays a large role in the temporal gap (of nearly thirty years) between the initial theoretical work and the fits of \citet{Gruen2018}. Early efforts such as \citet{Colombi1994} suggested that fitting log moments might be more tractable, but until now fits to the entire PDF have typically required an entire ensemble of cosmological simulations to estimate the covariance matrix.

Therefore, as an alternative to running a computationally expensive suite of simulations, we in this work derive analytical expressions for the variance and covariance of CIC probability measurements. To test our results, we empirically determine the variability of probability measurements in a Millennium Simulation galaxy catalog, and we show that our expressions accurately predict the variance and covariance of the measured galaxy PDF.

We structure the remainder of this work as follows: Section~\ref{sec:slicefield} defines \emph{slice fields} corresponding to probability bins; these slice fields form the foundation for the derivation. Section~\ref{sec:var} derives the variance of CIC measurements along with the expected error on the estimator for this variance; we then demonstrate the accuracy of our result by comparing it to simulation measurements of the CIC variance. Likewise, Section~\ref{sec:covar} derives the covariance between CIC measurements in different bins of probability; we again determine the expected error on the covariance estimator and demonstrate the accuracy of our result. Discussion and conclusions follow in Sections~\ref{sec:disc} and \ref{sec:concl}.

\section{The Slice Field}
\label{sec:slicefield}
Suppose we have a field of objects (such as a galaxies in a survey) contained in a number $N_c$ of non-overlapping cells positioned at $\mathbf{r}_1, \mathbf{r}_2, \ldots, \mathbf{r}_{N_c}$. If each cell contains a whole number $N_i = N(\mathbf{r}_i)$ of objects, it is straightforward to measure the probability distribution $\mathcal{P}(N)$ for the field. More generally, if we bin the measured numbers $N$ into non-overlapping bins $B_1, B_2, \ldots$, we can likewise measure the probability $\mathcal{P}(B)$ for any given bin $B$, recovering $\mathcal{P}(N)$ when the bins have unit width.

For a given bin $B$, we now define $\mathcal{S}_B$, the \emph{slice field} for $B$, such that $\mathcal{S}_B(\mathbf{r}_i) = 1$ if $N(\mathbf{r}_i) \in B$; otherwise $\mathcal{S}_B(\mathbf{r}_i) = 0$. This field $\mathcal{S}_B$ thus identifies the spatial location of a particular slice of the possible values of $N$, namely, those for which $N \in B$. Bins of unit width allow us to recover the original $N$-field of counts-in-cells (CIC) by summing over the slice fields:
\begin{equation}
N(\mathbf{r}_i) = \sum_N N\mathcal{S}_{\{N\}}(\mathbf{r}_i).
\end{equation}
Wider bins likewise recover a binned version of the original field.

Two simple slice-field properties will be useful in the sequel. First, because all slice-field values are either 0 or 1, all moments of any given slice field are equal to the probability of the associated bin:
\begin{equation}
\mathcal{P}(B) = \langle \mathcal{S}_B \rangle = \langle \mathcal{S}_B^{\:n} \rangle,
\label{eq:moms}
\end{equation}
for all natural numbers $n$. Second, if two bins $B_1$ and $B_2$ are disjoint, their corresponding slice fields must also be disjoint:
\begin{equation}
\mathcal{S}_{B_1} \cdot \mathcal{S}_{B_2} \equiv 0 \mbox{ if } B_1 \cap B_1 = \varnothing.
\label{eq:disjoint}
\end{equation}
The correlation function of the slice field is related to the sliced correlation functions as defined in \citet{NeyrinckSzapudi2018}; it is straightforward to recover the sliced correlation functions from these slice fields by a 1-point slice-averaging at one end.

\section{Counts-in-cells Variance}
\label{sec:var}

\subsection{Theoretical Prediction}
\label{sec:theovar}
We now turn to the problem of determining the error on the measured CIC probability in a way that accounts for correlations between neighboring cells. Given a bin $B$ of counts, and writing $\mathcal{S}$ for $\mathcal{S}_B$, we see from Equation~\ref{eq:moms} that the probability of that bin $\mathcal{P}(B) = \langle \mathcal{S} \rangle$. Furthermore, the variance of $\mathcal{S}$ is $\sigma^2_\mathcal{S} = \langle \mathcal{S}^2 \rangle - \langle \mathcal{S} \rangle^2 =  \mathcal{P}(B)\left(1 - \mathcal{P}(B) \right)$.

If the survey cells are uncorrelated, it follows that the variance of $\mathcal{P}(B)$ is given by
\begin{equation}
\sigma^2_{\mathcal{P}(B)} = \sigma^2_{\langle S \rangle} = \frac{\mathcal{P}(B)\left(1 - \mathcal{P}(B) \right)}{N_c}.
\label{eq:uncorrvar}
\end{equation}
(Note that this expression appears in \citet{Colombi1995} for $B = \{0\}$ with no correlation.) Equation~\ref{eq:uncorrvar} treats each cell as an independent measurement of the probability; correlations between cells will decrease the effective number of independent measurements, requiring modification of the expression.

Thus, to handle the case of correlated cells, we first consider the artificial case of an $\mathcal{S}$-field consisting of $n$ survey cells such that the correlation between any two of them is a fixed value $\xi$. Explicitly, if $i \neq j$, we have $\langle s_i s_j \rangle = (1+\xi) P^2$, where $P = \mathcal{P}(B)$ is the probability that $\mathcal{S} = 1$ in any given cell. (Note that $\xi$ is the correlation of $\mathcal{S}$, not of the counts-in-cells $N$.) If $s_i$ is the value of the $\mathcal{S}$ in the $i$th cell, we know that
\begin{equation}
\mathcal{P}(B) = \langle \mathcal{S} \rangle = \frac{1}{n} \sum_{i=1}^n s_i.
\label{eq:meanP}
\end{equation}
In Appendix~\ref{app:var} we show that  the variance $\sigma^2_{\mathcal{P}(B)}$ of this quantity is given by
\begin{equation}
\mathrm{Var}\left( \frac{1}{n} \sum_{i=1}^n s_i \right) = \frac{\left(1 - \left(1-(n-1)\xi\right)P\right)P}{n},
\label{eq:IHvar}
\end{equation}

However, this expression presupposes an equal degree of correlation $\xi$ between all $n$ cells. Obtaining a similar expression for the general case of varying $\xi$ is analytically intractable. However, we note that, in the inductive proof of Equation~\ref{eq:IHvar} (Appendix~\ref{app:var}), the term through which new instances of $\xi$ accumulate into the expression is $\langle s_i s_{n+1} \rangle$ (in Equation~\ref{eq:rhsindvar}). This term has precisely the form expected for a Monte Carlo volume average of a spatially-varying $\xi(r)$. With this motivation, we approximate $\xi$ in Equation~\ref{eq:IHvar} with the average slice-field correlation $\overline{\xi}_\mathcal{S}$, where the average is taken over all pairs $(\mathbf{r}_i, \mathbf{r}_j)$ of survey cells ($i \neq j$). (In Section~\ref{sec:vartest} we verify the validity of this approximation numerically.) Performing this substitution and writing $N_c$ (the number of cells) for $n$, we obtain for any counts-in-cells bin $B$,
\begin{equation}
\sigma^2_{\mathcal{P}(B)} = \frac{\mathcal{P}(B)\left(1-\mathcal{P}(B)\right)}{N_c} + \frac{(N_c - 1) \overline{\xi}_\mathcal{S}}{N_c} \mathcal{P}(B)^2,
\label{eq:corrvar}
\end{equation}
which reduces to Equation~\ref{eq:uncorrvar} in the uncorrelated case.

\subsection{Error on the Measured Variance}
\label{sec:varvar}
In order to test Equation~\ref{eq:corrvar}, we need multiple measurements of the probability $\mathcal{P}(B)$. It is possible to obtain such measurements from a single simulation by dividing it into multiple subvolumes, calculating $\mathcal{P}(B)$ in each subvolume, and then observing the variance of those measured $\mathcal{P}(B)$-values. However, a meaningful comparison between measurement and theory requires an estimate of the possible scatter in these observations of the variance. To this we now turn.

Suppose we have a set $\left\{ P_i \right\}$ of $N_m$ measurements of a probability $\mathcal{P}(B)$. Let us denote the measured variance of this set as $s_P^2$. Then according to a standard result (e.g., \citealp[eq.~12.35]{KendallStuart}, converting cumulants to moments),
\begin{equation}
\mathrm{Var}\left(s_p^2\right) = \frac{\mu_4 - \mu_2^2}{N_m} + \frac{2\mu_2^2}{N_m(N_m-1)},
\label{eq:vars2form}
\end{equation}
where $\mu_j$ denotes the $j$th central moment of the distribution of $P_i$-values. We can use the measured variance $s_P^2$ for $\mu_2$; however, we must perform an additional estimation of $\mu_4$. For simplicity, in calculating $\mu_4$ we will ignore the correlation between neighboring cells; we shall however mitigate the effect of this simplification by expressing our results in terms of $s_p^2$, which do include the effects of correlation.

Proceding to determine $\mu_4$, and employing $P$ as shorthand for $\mathcal{P}(B)$, it is straightforward to obtain the moment-generating function for the slice-field $\mathcal{S}$ corresponding to bin $B$:
\begin{equation}
M_\mathcal{S}(t) = 1 + \left(e^t - 1\right)P.
\end{equation}
Now, if each measurement $P_i$ was obtained by averaging the $\mathcal{S}$-values in $N_c$ cells (see Equation~\ref{eq:meanP}), then the moment-generating function for the probability is
\begin{equation}
M_\mathcal{P}(t) = \frac{1}{N_c} \left( 1 + \left(e^t - 1\right)P \right)^{N_c}.
\label{eq:MGFP}
\end{equation}

From this function (see Appendix~\ref{app:varvar} for details) we obtain the following expression for $\mu_4$:
\begin{equation}
\mu_4 = 3\left(s_p^2\right)^2 + \frac{s_p^2}{N_c}(1-6P+6P^2);
\label{eq:mu4}
\end{equation}
inserting this result into Equation~\ref{eq:vars2form} and simplifying, we obtain
\begin{equation}
\mathrm{Var}\left(s_p^2\right) = \frac{2\left(s_p^2\right)^2}{N_m - 1} + \frac{s_p^2(1-6P+6P^2)}{N_m N_c^2},
\label{eq:varvar}
\end{equation}
where $N_c$ is the number of survey cells in each measurement of $s_P^2$, and $N_m$ is the total number of measurements. Equation~\ref{eq:varvar} thus gives us the uncertainty on the measured value of $s_p^2$.

Note also that each probability measurement $P_i$ is a mean of $N_c$ values of the $\mathcal{S}$-field. Thus for reasonably large values of $N_c$, we can invoke the central limit theorem and treat the distribution of $P_i$-values as Gaussian, in which case $\mathrm{Var}\left(s_p^2\right) = 2\left(s_p^2\right)^2\!/(N_m - 1)$. This result is, of course, the limit of Equation~\ref{eq:varvar} for large $N_c$.

\subsection{Comparing Theory with Measurement}
\label{sec:vartest}
\begin{figure*}
    \leavevmode
    \epsfxsize=18cm
    \epsfbox{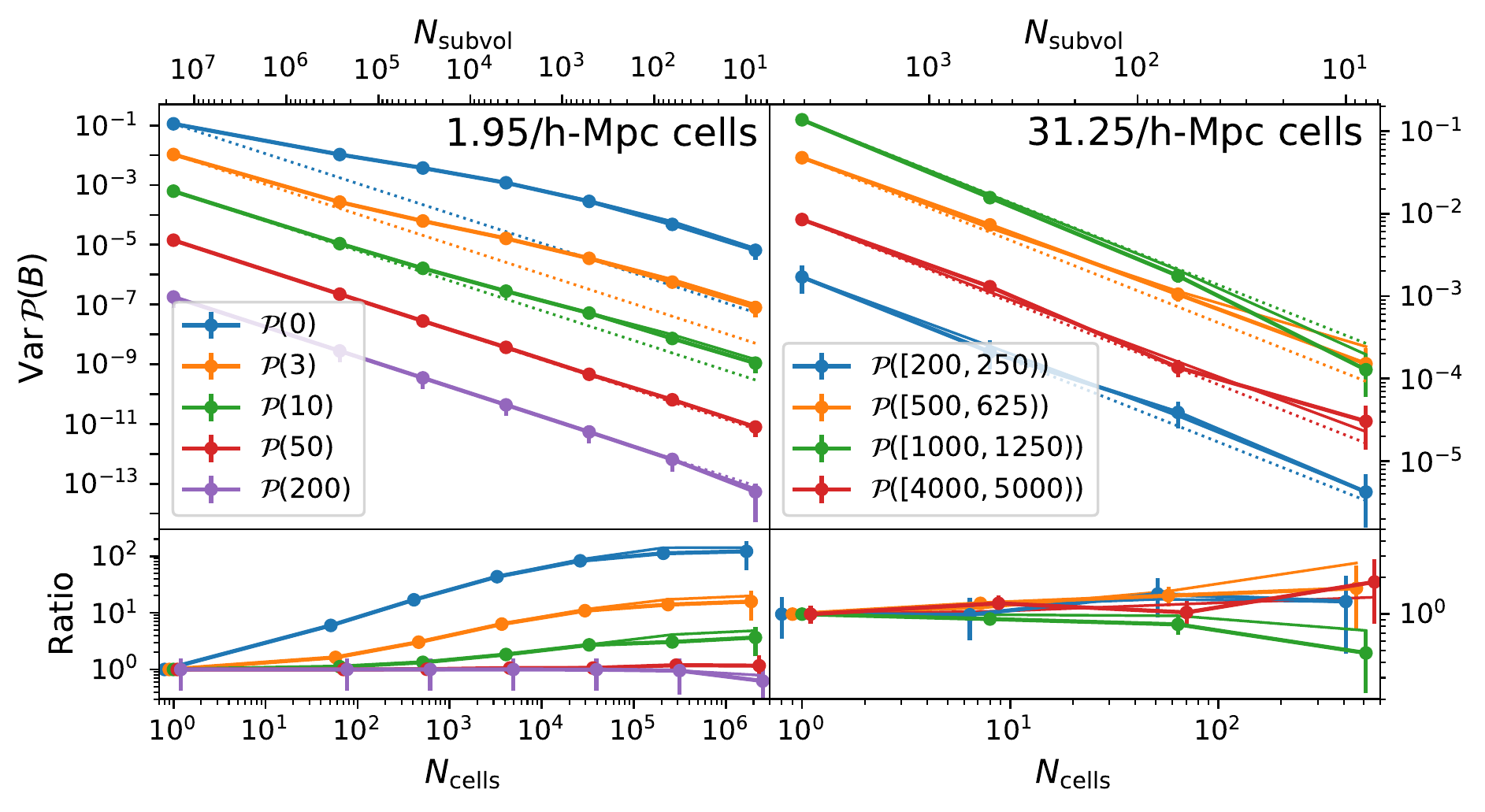}
    \caption{A comparison of the variance of CIC-probabilities predicted by Equation~\ref{eq:corrvar} with those measured in a mock galaxy survey catalog, in $1.95h^{-1}$-Mpc cubical cells (left panels) and $31.25h^{-1}$-Mpc cubical cells (right panels). The top axes show the number of subvolumes into which the survey is split  ($N_m$ in Equation~\ref{eq:varvar}), equal to the number of measurements of $\mathcal{P}(B)$; the bottom axes show the number of cells within each subvolume ($N_c$ in Equation~\ref{eq:varvar}). Thick lines (with error bars) show the empirical variance of the measurements of $\mathcal{P}(B)$ (one measurement for each subvolume, each subvolume containing $N_\mathrm{cells}$ cells); thin lines show the predicted variances from Equation~\ref{eq:corrvar}, where we determine the volume-averaged correlations as describe in the text. The thin dotted lines show the predicted variance for the uncorrelated case (Equation~\ref{eq:uncorrvar}). The bottom panels display the ratio of the true variance to that predicted in the absence of correlation; in some cases the correlations can increase the variance by two orders of magnitude. For clarity, the lower-panel curves have received a slight horizontal offset from each other.}
\label{fig:varprob}
\end{figure*}
 
We can now compare the predictions of Equation~\ref{eq:corrvar} with measured variances. To do so, we make use of the L-galaxies catalog\footnote{From the repository at \texttt{http://gavo.mpa-garching.mpg.de/\\Millennium/}} \citep{Bertone2007} from the Millennium Simulation \citep{Springel2005}, imposing a stellar mass cut of $M_\star \ge 10^9 \mathrm{M}_\odot$ to obtain a mock galaxy survey. We perform two tests, one with the survey volume divided into $256^3$ cubical cells (with side length $1.95h^{-1}$Mpc) and a second with the volume divided into $16^3$ cubical cells (with side length $31.25h^{-1}$Mpc).

We next split the survey into $N_m$ subvolumes, each consisting of $N_c$ survey cells (so that $N_m N_c = N_\mathrm{tot}$, the total number of cells in the survey). Given a CIC bin $B$, we determine the probability $\mathcal{P}(B)$ within each subvolume, thus obtaining an ensemble of $N_m$ measured probabilities. The variance of this ensemble gives us a measured value for $\sigma^2_\mathcal{P}(B)$, and this value will depend on the number of cells $N_c$ used in the measurement of each probability. These empirical variances appear as thick curves in Fig.~\ref{fig:varprob}. In this figure, the bottom axis shows $N_c$, the number of cells in each subvolume, and the top axis shows $N_m$, the number of subvolumes. (Note also that we choose unit bin widths for the $1.95h^{-1}$-Mpc cells and varying bin widths for the $31.25h^{-1}$-Mpc cells.) Equation~\ref{eq:varvar} gives us the error bars on these measurements, where we use for $s^2_p$ and $P$ the measured variances and probabilities.

The thin curves in Fig.~\ref{fig:varprob} show the predictions of Equation~\ref{eq:corrvar}. This prediction is not entirely \emph{a priori}, since it requires the (measured) probability values $\mathcal{P}(B)$ and the measured volume-averaged correlation $\overline{\xi}_\mathcal{S}$ of the corresponding slice field. To obtain the latter, we first measure the two-point correlation function $\xi_\mathcal{S}(r)$ of the appropriate slice field using a standard fast Fourier transform method; we then use Monte Carlo sampling of the slice field to obtain random pairs, using $\xi(r)$ to calculate their correlation and folding the result into the average; we continue the sampling process until the variation in the running average has subpercent effect on the predicted  $\sigma^2_\mathcal{P}(B)$.

The endpoints of the curves illustrate two extremes. The left-hand endpoints represent the situation in which each survey cell constitutes its own subvolume ($N_c=1$, and $N_m=256^3$ or $16^3$). In this case, each cell provides an estimate of $\mathcal{P}(B)$, and these estimates are either 0 or 1 (depending on whether the cell falls into that probability bin). In this case we expect the variance of these estimates to be large.

The right-hand endpoint of each curve represents the opposite situation of high $N_c$ and low $N_m$. At this endpoint, the survey is divided into 8 subvolumes ($N_c=256^3\!/8$ or $16^3\!/8$, and in both cases $N_m=8$). Here we have 8 measurements of $\mathcal{P}(B)$, and the variance of these measurements is small due to the large number of cells $N_c$ involved in the calculation of each one. On the other hand, since we have only 8 measurements of $\mathcal{P}(B)$, our estimate $s^2$ of the variance is less certain.

In both cases ($1.95h^{-1}$- and $31.25h^{-1}$-Mpc cells), we see from this figure that the predicted variance is in excellent agreement with the measured values. Furthermore, we see the expected trends: as the number of cells $N_c$ used for the measurement of $\mathcal{P}(B)$ increases, the variance in those measurements decreases; however, as the number of measurements of $\mathcal{P}(B)$ decreases, the uncertainty in the variance increases.

The top panels of Fig.~\ref{fig:varprob} also show, as thin dotted lines, the variance predicted under the assumption of no correlation between survey cells (Equation~\ref{eq:uncorrvar}), which is proportional to $1/N_c$; the bottom panels show the ratio between the correlated and uncorrelated cases. It is evident that inter-cell correlations can have a significant effect on the variance of $\mathcal{P}(B)$; in the case of $\mathcal{P}(0)$ for the $1.95h^{-1}$-Mpc cells, the difference is more than two orders of magnitude.

\section{Counts-in-cells Covariance}
\label{sec:covar}

\subsection{Theoretical Prediction}

The second issue one must consider in fitting models to CIC results is the covariance between different bins of counts. (It is clear that such covariance must exist, given that a survey cell falling into one bin is thereby excluded from all other bins.)  Thus we here derive an expression for the covariance $\sigma_{\mathcal{P}(B_1)\mathcal{P}(B_2)}$ of counts-in-cells in two (disjoint) bins $B_1$ and $B_2$. To simplify notation, we write $\mathcal{S}_1$, $\mathcal{S}_2$ for the slice fields of the two bins, and $P_1$, $P_2$ for $\mathcal{P}(B_1)$, $\mathcal{P}(B_2)$.

We begin by considering the case of two slice fields, both drawn from the same survey consisting of $n$ cells, and, as before, we initially assume that the cross-correlation between the two slice fields is a fixed, constant value $\xi_{12}$. In particular, for $i \neq j$ we let $s_{1i} = \mathcal{S}_1(\mathrm{r}_i)$ and $s_{2j} = \mathcal{S}_2(\mathrm{r}_j)$; then given this cross-correlation, we can say that the joint probability of $(s_{1i}=1,s_{2j}=1)$ is $\mathcal{P}(1,1) = P_1 P_2 (1 + \xi_{12})$. Furthermore, since  $\mathcal{S}_1(\mathrm{r}_i)\mathcal{S}_2(\mathrm{r}_j)$ vanishes unless $\mathcal{S}_1(\mathrm{r}_i) = \mathcal{S}_2(\mathrm{r}_j) = 1$, the expected value $\langle \mathcal{S}_1(\mathrm{r}_i)\mathcal{S}_2(\mathrm{r}_j)\rangle = \mathcal{P}(1,1)$.

Now $P_1$ is simply $(1/n)\sum s_{1i}$, and $P_2$ is $(1/n)\sum s_{2j}$. Given these relationships, we prove in Appendix~\ref{app:covar} the following statement (analogous to Equation~\ref{eq:IHvar}) concerning the covariance of the probabilities $P_1$ and $P_2$:
\begin{equation}
\mathrm{Cov}\left( \frac{1}{n} \sum_{i=1}^n s_{1i}\,,  \frac{1}{n} \sum_{j=1}^n s_{2j}\right) = \frac{-P_1 P_2}{n}\left(1 + (n-1)\xi_{12}\right).
\label{eq:IHcov}
\end{equation}
At this point we make an approximation analogous to that in Section~\ref{sec:theovar} by replacing the fixed $\xi_{12}$ with the average cross-correlation $\overline{\xi}_{\mathcal{S}_1\mathcal{S}_2}$ of the two slice fields. Again writing $N_c$ for $n$, we obtain, for any disjoint counts-in-cells bins $B_1$ and $B_2$,
\begin{equation}
\sigma_{\mathcal{P}(B_1)\mathcal{P}(B_2)} = \frac{-\mathcal{P}(B_1) \mathcal{P}(B_2)}{N_c} - \frac{n-1}{N_c} \overline{\xi}_{\mathcal{S}_1\mathcal{S}_2} \mathcal{P}(B_1) \mathcal{P}(B_2).
\label{eq:covar}
\end{equation}
We note that in the absence of cross-correlation, the covariance is negative since the bins are mutually exclusive.

\subsection{Error on the Measured Covariance}
As in Section~\ref{sec:var}, we now wish to compare Equation~\ref{eq:covar} with results measured from simulations, and thus we require an estimate for the uncertainty of the measured covariances.

Let us begin with two disjoint CIC bins $B_1$ and $B_2$; let us also suppose that we have two sets $\left\{ P_{1i} \right\}$ and $\left\{ P_{2i} \right\}$ of measurements of probabilities $\mathcal{P}(B_1)$ and $\mathcal{P}(B_2)$, each set consisting of $N_m$ measurements. We denote the covariance of these two sets of probability measurements as $S_{P_1 P_2}$.

The following expression \citep[p.~322]{KendallStuart} gives the variance of $S_{P_1 P_2}$ (where our $S_{P_1 P_2}$ corresponds to $k_{11}$ of \citeauthor{KendallStuart}):
\begin{equation}
\mathrm{Var}\left(S_{P_1 P_2}\right) = \frac{\mu_{22}}{N_m} + \frac{\mu_{02}\mu_{20}}{N_m(N_m-1)} - \frac{N_m-2}{N_m(N_m-1)}\mu_{11}^2,
\label{eq:KSvarcovar}
\end{equation}
where we have converted cumulants into moments, with $\mu_{rs}$ denoting the $r$th, $s$th product moments about the means of the random variables $P_1$, $P_2$. In calculating these moments we will ignore cross-correlations between nearby cells (as in Section~\ref{sec:varvar}), but we shall again seek to reduce the impact of this simplification by expressing our result in terms of $S_{P_1 P_2}$.

Let us employ $P_1$, $P_2$ as shorthand for $\mathcal{P}(B_1)$, $\mathcal{P}(B_2)$. Then the joint moment-generating function for the corresponding slice fields $\mathcal{S}_1$, $\mathcal{S}_2$ is
\begin{equation}
M_{\mathcal{S}_1 \mathcal{S}_2}(t_1,t_2) = 1 + \left( e^{t_1}-1 \right) P_1 + \left( e^{t_2}-1 \right) P_2.
\end{equation}
Furthermore (as in Section~\ref{sec:varvar}), each of the measurements $P_{1i}$, $P_{2i}$ is an average over the $\mathcal{S}_1$, $\mathcal{S}_2$-values in $N_c$ cells. Thus the joint moment-generating function for the measured probabilities is
\begin{equation}
M_{\mathcal{P}_1 \mathcal{P}_2}(t_1,t_2) = \frac{1}{N_c} \left(1 + \left( e^{t_1}-1 \right) P_1 + \left( e^{t_2}-1 \right) P_2\right)^{N_c}.
\label{eq:jntmgf}
\end{equation}
From this function, we calculate in Appendix~\ref{app:varcovar} the required joint cental moments of the $\mathcal{P}_1, \mathcal{P}_2$ distribution. We then apply these results to Equation~\ref{eq:KSvarcovar} and, using $S_{P_1 P_2} = -P_1 P_2 / N_c$ (Equation~\ref{eq:covar} with no cross-correlation), it eventually follows that
\begin{multline}
\mathrm{Var}\left(S_{P_1 P_2}\right) = \frac{1}{N_m-1} \left\{\rule{0pt}{12pt} 2\left(S_{P_1 P_2}\right)^2 \left(1 - \frac{3(N_m-1)}{N_c N_m}\right)\right.\\
 - \frac{S_{P_1 P_2}}{N_c^2 N_m} \left[\rule{0pt}{10pt}(1-P_2-P_2) (1 + N_m(N_c-1))\right.\\
 + \left.\left.(P_1 + P_2)(N_m - 1) \rule{0pt}{10pt}\right]\rule{0pt}{12pt}\right\}.
 \label{eq:varcovar}
\end{multline}
As with the variance in Section~\ref{sec:varvar}, we can note that for large values of $N_c$, the $P_{1i}$, $P_{2i}$ values quickly approach a joint Gaussian distribution; in this case we can take the limit of Equation~\ref{eq:varcovar} to obtain $\mathrm{Var}\left(S_{P_1 P_2}\right) = 2\left(S_{P_1 P_2}\right)^2 / (N_m -1)$.

\subsection{Comparing Theory with Measurement}
\label{sec:covartest}
\begin{figure*}
    \leavevmode
    \epsfxsize=18cm
    \epsfbox{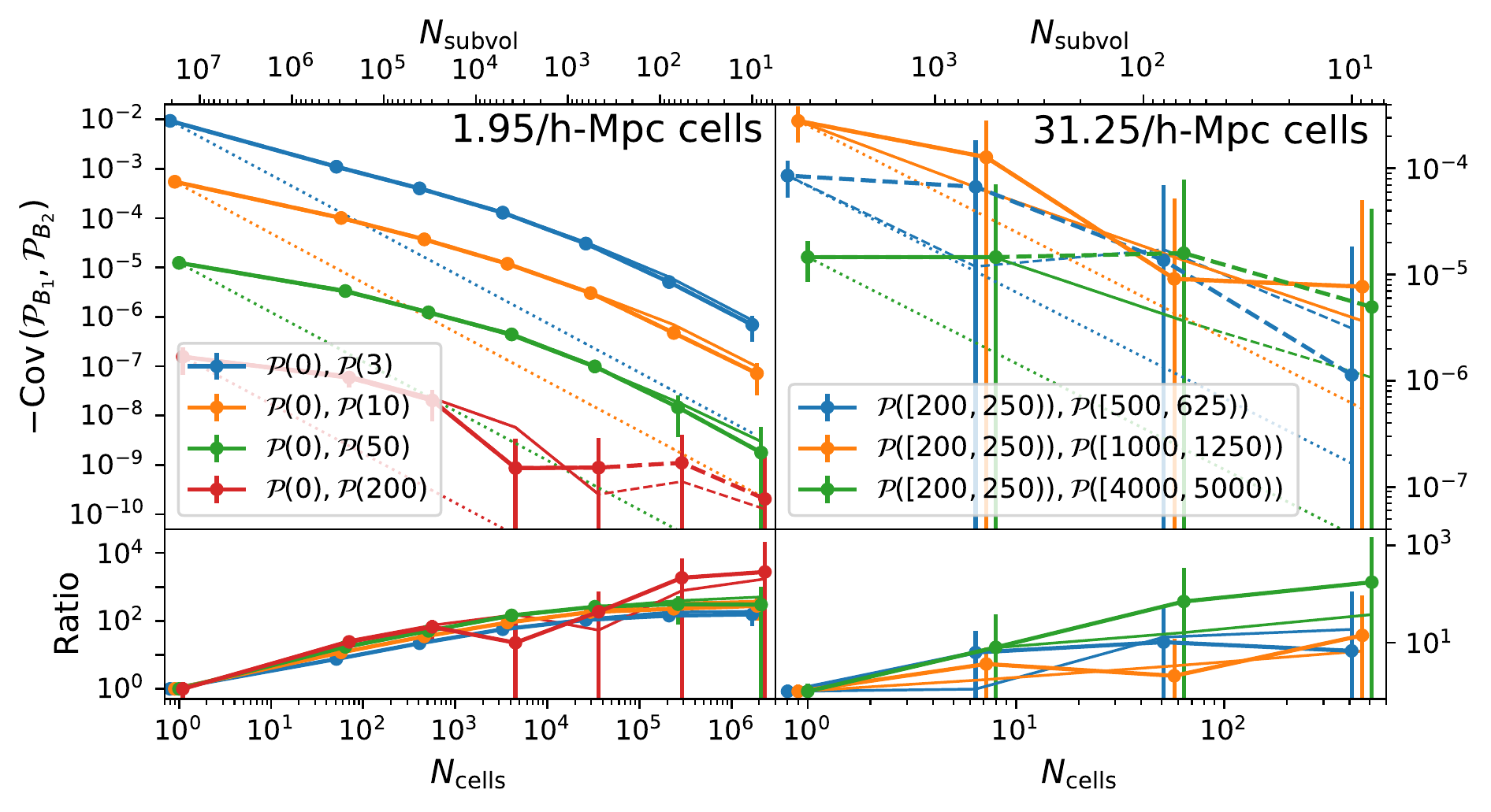}
    \caption{A comparison of the predicted covariance of CIC probabilities (Equation~\ref{eq:covar}) and the measured covariance in the same mock galaxy catalog as in Fig.~\ref{fig:varprob}.  The top axes show the number of subvolumes into which the survey is split ($N_m$ in Equation~\ref{eq:varcovar}), equal to the number of measurements of $\mathcal{P}(B_1)$ and $\mathcal{P}(B_2)$; the bottom axes show the number of cells in each subvolume ($N_c$ in Equation~\ref{eq:varcovar}). Thick lines (with error bars) show the negative of empirical covariance of the measurements of $\mathcal{P}(B_1)$ and $\mathcal{P}(B_2)$ (i.e., $N_\mathrm{subvol}$ measurements, each involving $N_\mathrm{cells}$ survey cells); thin lines show the negative of the predicted covariance from Equation~\ref{eq:covar}, using volume-averaged cross-correlations determined as in the text. The thin dotted lines show the negative of the predicted covariance for the case of no cross-correlation. Dashed lines (thick and thin) indicate positive (rather than negative) values for the covariance. The bottom panels display the absolute ratio of the true covariances to those predicted assuming no cross-correlation. For clarity, the curves have received a slight horizontal offset from each other.}
\label{fig:covarprob}
\end{figure*}

As in Section~\ref{sec:vartest}, we proceed to compare the predictions of Equation~\ref{eq:corrvar} with measured covariances; we use the same mock survey, with the same cell sizes ($1.95h^{-1}$ and $31.25h^{-1}$Mpc), as in Section~\ref{sec:vartest}. Again we split each survey into $N_m$ subvolumes, with each subvolume consisting of $N_c$ survey cells.

In this case we choose two CIC bins $B_1$ and $B_2$ and empirically determine the probabilities $\mathcal{P}(B_1)$ and $\mathcal{P}(B_1)$ within each subvolume; the result is an ensemble of $N_m$ measured probabilities $P_{1i}$ in $B_1$ and a corresponding ensemble of $N_m$ measured probabilities $P_{2i}$ within $B_2$. The covariance $S_{P_1 P_2}$ of these two sets of measurements is our estimate for $\mathrm{Cov}(\mathcal{P}(B_1)\mathcal{P}(B_2))$, and this value will depend on the number of cells $N_c$ used in the measurement of the probabilities. As before, we use unit bin widths with the $1.95h^{-1}$-Mpc cells and varying bin widths with the $31.25h^{-1}$-Mpc cells.

These empirical covariances appear as thick curves in Fig.~\ref{fig:covarprob}. Since in these cases the covariances are typically negative, we plot $-\mathrm{Cov}$ with solid lines (using dashed lines to indicate positive covariances). Equation~\ref{eq:varcovar} provides the error bars for these measurements, where we use for $S_{P_1 P_2}$, $P_1$, and $P_2$ the measured covariances and probabilities.

The thin curves in Fig.~\ref{fig:covarprob} show the predictions of Equation~\ref{eq:covar}. For the volume-averaged correlation $\overline{\xi}_{\mathcal{S}_1\mathcal{S}_2}$ of the slice fields for the two bins, we first  empirically calculate the two-point cross-correlation function $\xi_{\mathcal{S}_1\mathcal{S}_2}(r)$ of the two slice fields with FFT methods. We then, as in Section~\ref{sec:vartest}, sample the slice fields to obtain random pairs of positions within the survey volume, determine the cross-correlation of those positions from $\xi_{\mathcal{S}_1\mathcal{S}_2}(r)$, and fold the result into a running average, terminating the sampling process once the variation in the running average has subpercent effect on the predicted $\sigma_{\mathcal{P}(B_1)\mathcal{P}(B_2)}$.

Once again, the agreement between prediction and measurement is excellent, although the large error bars at $31h^{-1}$-Mpc cells mean that most of the measurements yield only upper limits of the absolute value. We also observe the same (expected) trends as in Fig.~\ref{fig:varprob}: the covariance of the CIC measurements decreases as the number of cells $N_c$ in each measurement increases; likewise the error bars on the covariance increase as the number $N_m$ of measurements decreases.

The thin dotted lines in Fig.~\ref{fig:covarprob} show the predicted covariance in the case of no cross-correlation, and the lower panels show the ratio between the cross-correlated and non-cross-correlated results. It is again clear that cross-correlations among survey cells in different probability bins can increase the covariance by multiple orders of magnitude.

\section{Discussion}
\label{sec:disc}
\begin{figure*}
    \leavevmode
    \epsfxsize=18cm
    \epsfbox{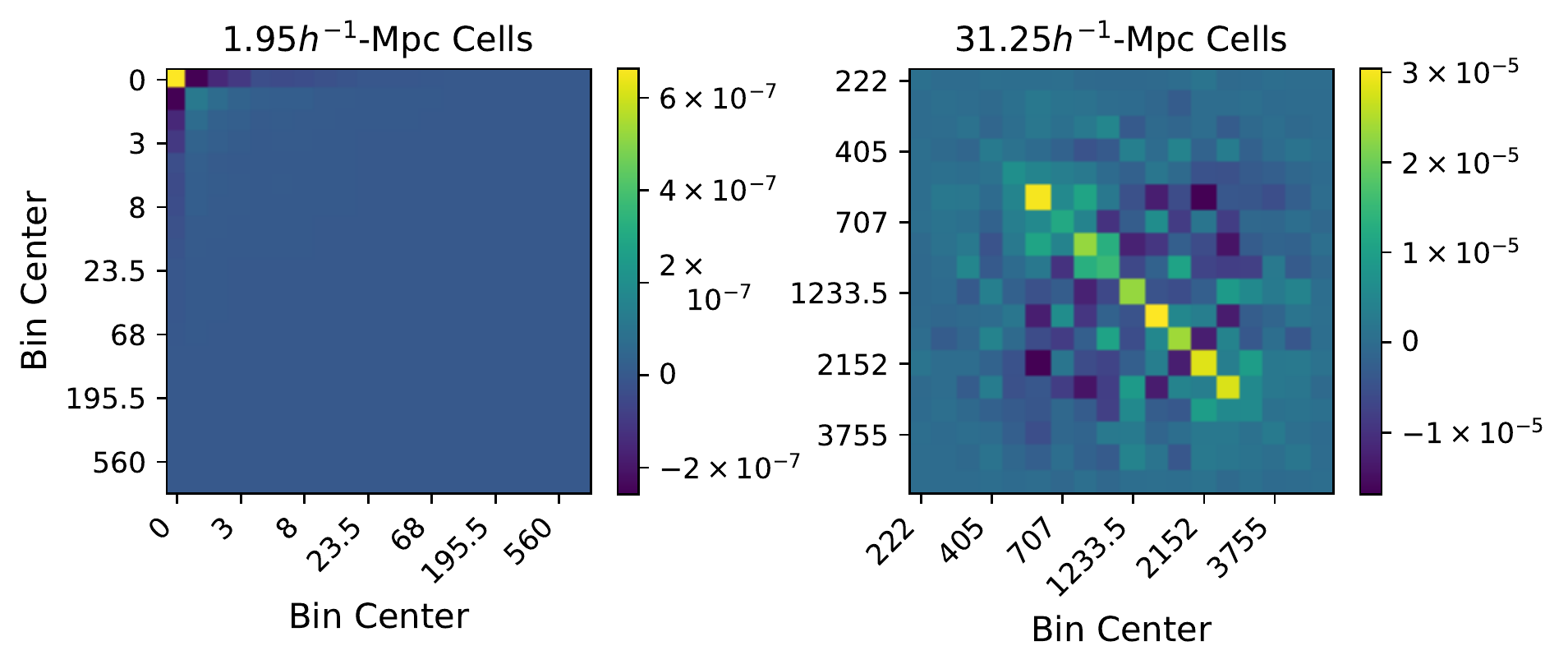}
    \caption{Covariance matrices for CIC probabilities in a mock galaxy survey (described in the text), using logarithmically spaced probability bins and measured (cross-) correlation functions.}
\label{fig:covmatr}
\end{figure*}

Equations~\ref{eq:corrvar} and \ref{eq:covar} allow us to calculate the covariance matrices for the mock surveys in Sections~\ref{sec:vartest} and \ref{sec:covartest} (though the calculation requires empirical determination of the average correlations $\overline{\xi}_\mathcal{S}$ and cross-correlations $\overline{\xi}_{\mathcal{S}_1 \mathcal{S}_2}$). We here perform one such calculation.

For our CIC bins, we start with 20 logarithmically-spaced bins, which we then combine as necessary to insure that no bin contains fewer than three survey cells; we end up with 20 and 18 bins for the $1.95h^{-1}$-Mpc and $31.25h^{-1}$-Mpc cases, respectively. Since we now use the entire survey to calculate the (co-)variances, we set $N_c = N_\mathrm{tot}$, the total number of survey cells, in Equations~\ref{eq:corrvar} and \ref{eq:covar}. Fig.~\ref{fig:covmatr} displays the resulting covariance matrices.

\begin{figure*}
    \leavevmode
    \epsfxsize=18cm
    \epsfbox{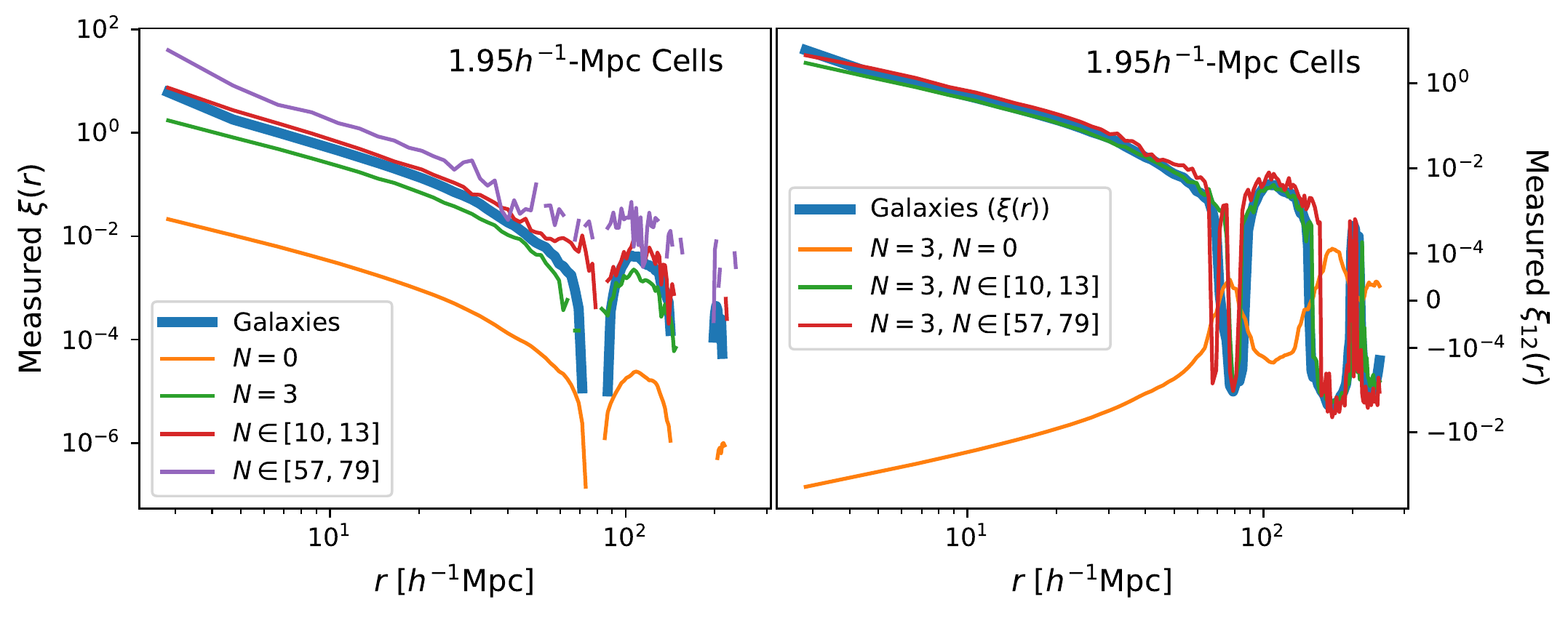}
    \caption{Left-hand panel: measured correlation function $\xi(r)$ for the slice fields of four CIC bins, compared to the galaxy correlation function, in our mock galaxy survey. Right-hand panel: measured cross-correlation function $\xi_{12}(r)$ between the $N = 3$ slice field and three other slice fields, again compared to the galaxy correlation function. To first order, the two-point (cross-) correlation functions for the slice fields seem to differ from the galaxy correlation function by a simple multiplicative factor.}
\label{fig:xis}
\end{figure*}
We note the following concerning these matrices. First, the two cases display significant structural differences. At $\sim 30h^{-1}$Mpc (right-hand panel), the covariance matrix is approximately diagonal (albeit with significant noise), indicating that at these scales the CIC galaxy distribution is approaching a Gaussian limit. However, at $\sim 2h^{-1}$Mpc (left-hand panel), the covariance is dominated by the $N=0$ cells, which occupy over 85 per cent of the survey volume. Furthermore, we have already noted (immediately following Equation~\ref{eq:covar}) that, in the absence of cross-correlations, the covariance between probability bins will be negative; we see this behavior in the left-hand panel of Fig.~\ref{fig:covmatr} near $N=0$. In this case, the negative covariance induced by mutual exclusivity is exacerbated by the fact that the empty cells aggregate into large voids and thus are negatively cross-correlated with $N>1$ (see right-hand panel Fig.~\ref{fig:xis}). Other than these features, the covariance matrix at this smaller scale is quite smooth.

The second observation is that, in consequence, to assume diagonal covariance matrices is a reasonable approximation on scales $\ga 30h^{-1}$Mpc. This fact is also evident in the right-hand panel of Fig.~\ref{fig:varprob}, which makes it clear that at such scales the  intercellular correlations have only a minor effect on the variance of $\mathcal{P}(B)$. \citet{ReppSDSSLin} are therefore justified in ignoring these correlations when fitting $\sigma_8$ and linear bias to CIC measurements from the Sloan Main Galaxy Sample. However, extraction of information from the galaxy PDF at smaller scales will need to take these correlations (and cross-correlations) into account.

The third observation is that the calculated covariance matrix is only as good as the measurement of the (cross-) correlation functions of the slice fields. It is this fact which is responsible for the noise in the right-hand panel of Fig.~\ref{fig:covmatr}, since at these scales we have only $16^3$ cells in our survey (and thus many fewer in each probability bin). As a result, we expect the measurement of $\overline{\xi}$ to be quite noisy, as we in fact see. Even at small scales, the measurement of $\xi(r)$, and thus of $\overline{\xi}$ becomes quite noisy for the low-probability, high-density bins (Fig.~\ref{fig:xis}). Thus, it would be helpful to have a theory for the slice-field correlation functions.

Such a theory seems to be feasible, given that the slice-field correlation and cross-correlation functions appear to differ from the galaxy correlation function by a multiplicative bias, at least to first order (Fig.~\ref{fig:xis}). Indeed, the slice fields pick out specific density contours in a manner analogous to the way in which galaxies preferentially trace regions of high matter-density; thus one might expect a bias analogous to the Kaiser bias of galaxies \citep{Kaiser1984}. Fitting these bias parameters to the measured correlations could therefore eliminate much of the noise from the measurements of the various $\overline{\xi}$-values.

Finally, we note that the correlations of the slice field represent a further generalization of the sliced correlation functions introduced by \citet{NeyrinckSzapudi2018}, which isolate the correlation of one particular density with the entire field. Thus they are similar to marked correlation functions and power spectra, which promise to enhance, e.g., the detection of neutrino signatures \citep{Massara2020,Philcox2020} in galaxy surveys. Indeed, since slice-field correlations are two-point functions -- whereas marked correlations are inherently higher-order (densities at two points plus spatially-varying marks) -- it is possible that slice-field correlations will prove more tractable than marked correlations, without sacrificing information content.

\section{Conclusions}
\label{sec:concl}
Since counts-in-cells (CIC) probabilities contain significant information not included in the galaxy power spectrum, it is important to develop the theoretical machinery for fitting cosmological models to CIC measurements from galaxy surveys. One of the key ingredients in performing such fits is an understanding of the variance of CIC-measurements within a given probability bin, as well as of the covariance of those measurements between different probability bins. We have here derived expressions for both of these quantities.

In order to derive these expressions, we first define the \emph{slice field} $\mathcal{S}_B$ for a given bin $B$, such that $\mathcal{S}_B = 1$ if $N$ (the number of galaxies within a survey cell) falls within $B$; otherwise $\mathcal{S}_B = 0$. Using these fields we derive Equation~\ref{eq:corrvar} for the variance $\mathrm{Var}(\mathcal{P}(B))$ of measurements of a given probability bin, and we derive Equation~\ref{eq:covar} for the covariance $\mathrm{Cov}(\mathcal{P}(B_1),\mathcal{P}(B_2))$ of the measurements in two distinct probability bins. These expressions depend on the probabilities, the number of cells from which the probabilities are determined, and the volume-averaged (cross-) correlation of the corresponding slice fields. Conceptually, the degree of correlation affects the result by reducing the effective number of cells involved in the probability calculation.

To test Equations~\ref{eq:corrvar} and \ref{eq:covar} we turn to a mock galaxy survey from the Millennium Simulation; by dividing the survey into multiple subvolumes we can measure the probability within each subvolume and thus empirically determine the variance/covariance of the $\mathcal{P}(B)$ measurements. Furthermore, a meaningful comparison to the predicted (co\nobreakdash-)variances requires an estimate of the scatter in the measurement of those (co\nobreakdash-)variances. Taking that scatter into account, we find that Equations~\ref{eq:corrvar} and \ref{eq:covar} accurately predict the variance and covariance of CIC measurements (Figs.~\ref{fig:varprob} and \ref{fig:covarprob}).

We further conclude that at large scales ($\sim 30h^{-1}$Mpc) the correlation between neighboring cells has a negligible impact on the variance, whereas at small scales ($\sim 2h^{-1}$Mpc) the correlations can increase the variance by orders of magnitude.

In summary, two of the tools necessary for wider cosmological utilization of counts-in-cells are the ability to determine the variance and covariance of the CIC probabilities. These tools are now available.

\section*{Acknowledgements}
The Millennium Simulation data bases used in this work and the web application providing online access to them were constructed as part of the activities of the German Astrophysical Virtual Observatory (GAVO). This work was supported by NASA Headquarters under the NASA Earth and Space Science Fellowship program -- ``Grant 80NSSC18K1081'' -- and AR gratefully acknowledges the support. IS acknowledges support from National Science Foundation (NSF) award 1616974. 

\section*{Data Availability}
The data underlying this article are available in the Virgo-Millennium Database (maintained by the German Astrophysical Virtual Observatory) at http://gavo.mpa-garching.mpg.de/Millennium.

\appendix
\section{Derivations}

\subsection{Proof of Equation~\ref{eq:IHvar}}
\label{app:var}
Equation~\ref{eq:IHvar} claims that the variance of $\mathcal{P}(B) = (1/n)\sum s_i$ is
\begin{equation}
\mathrm{Var}\left( \frac{1}{n} \sum_{i=1}^n s_i \right) = \frac{\left(1 - \left(1-(n-1)\xi\right)P\right)P}{n}.
\tag{\ref{eq:IHvar}}
\end{equation}
Recall that we are assuming the correlation between any two of the slice-field values $s_1, s_2, \ldots, s_n$ is a fixed quantity $\xi$. Thus if $i \neq j$, it is the case that $\langle s_i s_j \rangle = (1+\xi) P^2$, where $P = \mathcal{P}(B)$ is the probability that $\mathcal{S} = 1$ in any given cell. We also recall that $\xi$ is the correlation function of the slice field $\mathcal{S}$, not the of the counts-in-cells $N$.

The proof of Equation~\ref{eq:IHvar} proceeds by induction. For $n=1$, the right-hand side of Equation~\ref{eq:IHvar} is simply $P - P^2 = \langle \mathcal{S}^2 \rangle - \langle S \rangle^2$ (by Equation~\ref{eq:moms}), which is of course the variance of $\mathcal{S}$. On the other hand, for an arbitrary $n$, we have
\begin{gather}
\mathrm{Var}\left( \frac{1}{n+1} \sum_{i=1}^{n+1} s_i \right) = \frac{1}{(n+1)^2} \left\langle \left( \sum_{i=1}^{n+1} s_i - (n+1)P \right)^2 \right\rangle\hfill\\
  \quad = \frac{1}{(n+1)^2} \left\langle \left(\left( \sum_{i=1}^n s_i  - nP\right)  +  \left(s_{i+1}- P \right)\right)^2 \right\rangle\\
  \begin{split}
  \quad = \frac{1}{(n+1)^2} & \left\{ n^2\left\langle \left( \frac{1}{n}\sum_{i=1}^n s_i - P\right)^2\right\rangle\right.  \\
    & + 2\left\langle\left(\sum_{i=1}^n s_i - nP\right)\left(s_{n+1}-P\right)\right\rangle\\
    & + \left\langle\left(s_{n+1} - P\right)^2\right\rangle \left.\rule{0pt}{15pt}\right\}
   \end{split}\label{eq:3terms}
\end{gather}
Equation~\ref{eq:3terms} contains three expectation values. If Equation~\ref{eq:IHvar} holds for $n$, the first expectation value equals $\left(1 - \left(1-(n-1)\xi\right)P\right)P/n$. The third is simply the variance of $\mathcal{S}$, or $P-P^2$. And for the second, we have
\begin{gather}
   \begin{split}
   &\left\langle\left(\sum_{i=1}^n s_i - nP\right)\left(s_{n+1}-P\right)\right\rangle\\
   &\quad = \sum_{i=1}^n \left\langle s_i s_{n+1} \right\rangle - P \sum_{i=1}^n \langle s_i \rangle - nP\langle s_{i+1} \rangle + nP^2
   \label{eq:rhsindvar}
   \end{split}\\
   \quad= nP^2(1 + \xi) - nP^2 = nP^2\xi.
\end{gather}
 
Substituting these expressions for the expectation values into Equation~\ref{eq:3terms} and simplifying, we obtain
\begin{equation}
\mathrm{Var}\left( \frac{1}{n+1} \sum_{i=1}^{n+1} s_i \right) = \frac{\left(1-P\left(1-n\xi\right)\right)}{n+1},
\end{equation}
completing the induction on Equation~\ref{eq:IHvar}.

\subsection{Moments of of the Distribution of Measurements of $\mathcal{P}(B)$}
\label{app:varvar}
If $B$ is a bin in which we measure the probability $\mathcal{P}(B)$, let us suppose that $\{ P_i \}$ is a set of $N_m$ measurements of $\mathcal{P}(B)$. Our goal is to determine the fourth central moment $\mu_4$ of the distribution of the measurements $\{ P_i \}$ of $\mathcal{P}(B)$. We begin with Equation~\ref{eq:MGFP} which gives the moment-generating function for this distribution:
\begin{equation}
M_\mathcal{P}(t) = \frac{1}{N_c} \left( 1 + \left(e^t - 1\right)P \right)^{N_c},
\tag{\ref{eq:MGFP}}
\end{equation}
from which we can determine the moments of the distribution of measured $\mathcal{P}(B)$ values.

Using $P$ as a shorthand for $\mathcal{P}(B)$, we obtain the following moments:
\begin{gather}
\label{eq:mP1} \left\langle P \right\rangle = P\\
\label{eq:mP2} \left\langle P^2 \right\rangle = \frac{P}{N_c} + \frac{(N_c-1)P^2}{N_c}\\
\left\langle P^3 \right\rangle = \frac{P}{N_c^2} + \frac{3(N_c-1)P^2}{N_c^2} + \frac{(N_c-2)(N_c-1)P^3}{N_c^2}\\
\begin{split}
\left\langle P^4 \right\rangle = \frac{P}{N_c^3} + \frac{7(N_c-1)P^2}{N_c^3} + \frac{6(N_c-2)(N_c-1)P^3}{N_c^3}\\
 + \frac{(N-3)(N-2)(N-1)P^4}{N_c^3}.
\end{split}
\end{gather}

Thus the fourth central moment of the distribution of measured values for $\mathcal{P}(B)$ is
\begin{align}
\mu_4 & = \left\langle P^4 \right\rangle - 4\left\langle P^3 \right\rangle\left\langle P \right\rangle +6\left\langle P^2 \right\rangle\left\langle P \right\rangle^2 -3\left\langle P \right\rangle^4\\
& = \frac{3P^2}{N_c^2}(P-1)^2 + \frac{P(1-P)}{N_c^3}(1-6P+6P^2)\\
& = 3\left(s_p^2\right)^2 + \frac{s_p^2}{N_c}(1-6P+6P^2),
\end{align}
where the final step follows from $s_P^2 = P(1-P)/N_c$ (Equation~\ref{eq:uncorrvar}).

\subsection{Proof of Equation~\ref{eq:IHcov}}
\label{app:covar}
Equation~\ref{eq:IHcov} claims that the variance of the probabilities $P_1$ and $P_2$ is given by
\begin{equation}
\mathrm{Cov}\left( \frac{1}{n} \sum_{i=1}^n s_{1i}\,,  \frac{1}{n} \sum_{j=1}^n s_{2j}\right) = \frac{-P_1 P_2}{n}\left(1 + (n-1)\xi_{12}\right).
\tag{\ref{eq:IHcov}}
\end{equation}
As in Appendix~\ref{app:var}, we prove the claim using induction.

To establish Equation~\ref{eq:IHcov} for $n=1$, we first note that, because the slice fields are disjoint (Equation~\ref{eq:disjoint}), it cannot be the case that $\mathcal{S}_1 = \mathcal{S}_2 = 1$ in this single survey cell. The possibilities therefore are $\mathcal{S}_1 = 1, \mathcal{S}_2 = 0$ (with probability $P_1$), $\mathcal{S}_1 = 0, \mathcal{S}_2 = 1$ (with probability $P_1$), and $\mathcal{S}_1 = \mathcal{S}_2 = 0$ (with probability $1-P_1-P_2$). Hence the left-hand side of Equation~\ref{eq:IHcov} is
\begin{align}
\mathrm{Cov}(\mathcal{S}_1, \mathcal{S}_2) & = \left\langle(\mathcal{S}_1 - P_1)(\mathcal{S}_2 - P_2) \right\rangle\\
\begin{split}
   & = (1 - P_1)(-P_2)P_1 + (-P_1)(1 - P_2)P_2\\
   &\qquad + (P_1P_2)(1-P_1-P_2)
\end{split}\\
   & = -P_1P_2,
\end{align}
thus establishing the prescription for $n = 1$.

Now assuming the equation holds for a given $n$, we can write
\begin{align}
\mathrm{Cov}&\left(\frac{1}{n+1} \sum_{i=1}^{n+1} s_{1i}\,,  \frac{1}{n+1} \sum_{j=1}^{n+1} s_{2j}\right)\hspace{2cm}\\
\begin{split}
    &= \frac{1}{(n+1)^2} \left\langle\left(\sum_{i=1}^{n+1}s_{1i} - (n+1)P_1 \right)\right.\\
    & \qquad\qquad\qquad\qquad \left.\times\left(\sum_{j=1}^{n+1}s_{2j} - (n+1)P_2 \right)\right\rangle
\end{split}\\
\begin{split}
    &= \frac{1}{(n+1)^2} \left\{\left\langle\left(\sum_{i=1}^n s_{1i} - nP_1 \right)\left(\sum_{j=1}^n s_{2j} - nP_2 \right)\right\rangle\right.\\
    & \qquad\qquad+ \left\langle\left(\sum_{i=1}^n s_{1i} - nP_1 \right)\left(s_{2(n+1)} - P_2 \right)\right\rangle\\
    & \qquad\qquad+ \left\langle\left(s_{1(n+1)} - P_1 \right)\left(\sum_{j=1}^n s_{2j} - nP_2 \right)\right\rangle\\
    & \qquad\qquad+ \left\langle\left(s_{1(n+1)} - P_1 \right)\left(s_{2(n+1)} - P_2 \right)\right\rangle\left.\rule{0pt}{15pt}\right\}
\end{split}\label{eq:4evs}
\end{align}
Equation~\ref{eq:4evs} contains four expectation values which we now evaluate. By our inductive hypothesis, the first is $P_1 P_2n(1+(n-1)\xi_{12})$. The second becomes
\begin{align}
\sum_{i=1}^{n} & \langle s_{1i} s_{2(n+1)} \rangle - P_2 \sum_{i=1}^{n} \langle s_{1i} \rangle - nP_1 \langle s_{2(n+1)} \rangle + nP_1P_2\\
& = nP_1P_2 \xi_{12}
\end{align}
by recalling that for $i \neq j$, $\langle s_{1i}  s_{2j}\rangle = P_1 P_2 (1 + \xi_{12})$ and that $\langle s_{1i} \rangle = P1$, etc. The third expectation value becomes the same quantity by symmetry. Finally, we recall that $\langle s_{1(n+1)} s_{2(n+1)} \rangle = 0$ because the slice fields are disjoint, and thus the fourth expectation value becomes $-P_1P_2$.

Inserting these results into Equation~\ref{eq:4evs} and simplifying, we obtain
\begin{multline}
\mathrm{Cov}\left(\frac{1}{n+1} \sum_{i=1}^{n+1} s_{1i}\,,  \frac{1}{n+1} \sum_{j=1}^{n+1} s_{2j}\right)\\
    = \frac{-P_1 P_2}{n+1} (1 + n\xi_{12}),
\end{multline}
thus establishing Equation~\ref{eq:IHcov} for all natural numbers $n$.

\subsection{Moments of of the Joint Distribution of Measurements of $\mathcal{P}(B_1, B_2)$}
\label{app:varcovar}
If $B_1, B_2$ are disjoint CIC bins, let us consider two sets of measurements $\{ P_{1i}\}, \{P_{2i}\}$ of the probabilities $\mathcal{P}(B_1)$ and $\mathcal{P}(B_2)$, respectively, each consisting of $N_m$ measurements. We want to determine the central moments of the joint distribution of the measurements $\{ P_{1i}\}$ of $\mathcal{P}(B_1)$ and $\{P_{2i}\}$ of $\mathcal{P}(B_2)$. We start with Equation~\ref{eq:jntmgf} which (using $P_1$ and $P_2$ as shorthand for $\mathcal{P}(B_1)$ and $\mathcal{P}(B_2)$, respectively) gives the joint moment-generating function for this distribution:
\begin{equation}
M_{\mathcal{P}_1 \mathcal{P}_2}(t_1,t_2) = \frac{1}{N_c} \left(1 + \left( e^{t_1}-1 \right) P_1 + \left( e^{t_2}-1 \right) P_2\right)^{N_c}.
\tag{\ref{eq:jntmgf}}
\end{equation}
Proceding to calculate the joint moments by differentiating this expression, we obtain:
\begin{gather}
\left\langle P_1 P_2 \right\rangle = \frac{N_c-1}{N_c}P_1 P_2\\
\left\langle P_1^2 P_2 \right\rangle = \frac{N_c-1}{N_c^2}P_1 P_2 + \frac{(N_c-1)(N_c-2)}{N_c^2}P_1^2 P_2\\
\left\langle P_1 P_2^2 \right\rangle = \frac{N_c-1}{N_c^2}P_1 P_2 + \frac{(N_c-1)(N_c-2)}{N_c^2}P_1 P_2^2\\
\begin{split}
\left\langle P_1^2 P_2^2 \right\rangle =  &\frac{(N_c-1)(N_c-2)(N_c-3)}{N_c^3}P_1^2 P_2^2\\
 & + \frac{(N_c-1)(N_c-2)}{N_c^3}(P_1^2 P_2 + P_1 P_2^2)\\
 & + \frac{(N_c-1)}{N_c^3} P_1^2 P_2^2.
\end{split}
\end{gather}
From these results (along with Equations~\ref{eq:mP1} and \ref{eq:mP2}) we can obtain the required joint central moments of the $\mathcal{P}_1,\mathcal{P}_2$ distribution:
\begin{gather}
\mu_{20} = \left\langle P_1^2 \right\rangle - \left\langle P_1 \right\rangle^2 = \frac{P_1(1- P_1)}{N_c}\\
\mu_{02} = \left\langle P_2^2 \right\rangle - \left\langle P_2 \right\rangle^2 = \frac{P_2(1- P_2)}{N_c}\\
\mu_{11} = \left\langle P_1 P_2 \right\rangle - \left\langle P_1 \right\rangle \left\langle P_2 \right\rangle = -\frac{P_1 P_2}{N_c}
\end{gather}
\begin{align}
\begin{split}
\mu_{22} & = \left\langle P_1^2 P_2^2 \right\rangle - 2\left\langle P_1^2 P_2\right\rangle  \left\langle P_2 \right\rangle - 2\left\langle P_1 P_2^2\right\rangle  \left\langle P_1 \right\rangle\\
  &\qquad + 4\left\langle P_1 P_2\right\rangle \left\langle P_1 \right\rangle \left\langle P_2 \right\rangle + \left\langle P_1^2 \right\rangle \left\langle P_2\right\rangle^2 + \left\langle P_2^2 \right\rangle \left\langle P_1 \right\rangle^2\\
  &\qquad - 3\left\langle P_1 \right\rangle^2 \left\langle P_2 \right\rangle^2
\end{split}\\
& = \frac{P_1 P_2}{N_c^3} \left( (N_c-1)- (N_c-2)(P_1 + P_2 - 3P_1 P_2)\right).
\end{align}

\bibliographystyle{astron}
\bibliography{Thesis_Proposal}

\label{lastpage}

\end{document}